\documentclass[]{aa}  

\usepackage{graphicx}
\usepackage{txfonts}
\usepackage{natbib}
\begin{document} 

\title{Two epoch spectro-imagery of FS Tau B outflow system \thanks{This work is  based on observations conducted with the 6 m telescope of the Special Astrophysical Observatory of the Russian Academy of Sciences carried out with the financial support of the Ministry of Science and Higher Education of the Russian Federation.
   }} 

\titlerunning{FS Tau B}

   \author{T.A. Movsessian
          \inst{1}
          \and
          T.Yu. Magakian
          \inst{1}
          \and  
          A.N. Burenkov
          \inst{2}
          }
          
   \institute{Byurakan Astrophysical Observatory NAS Armenia, Byurakan, Aragatsotn prov., 0213,
              Armenia\\
                \email{tigmov@bao.sci.am; tigmag@sci.am}
          \and
              Special Astrophysical Observatory,
             N.Arkhyz, Karachaevo-Cherkesia, 369167 Russia\\
             \email{ban@sao.ru}
                        }

   \date{Received ...; accepted ...}

 
  \abstract
   {Herbig-Haro (HH) flows exhibit a large variety of morphological and kinematical structures such as bow shocks, Mach disks, and deflection shocks. Both proper motion (PM) and radial velocity investigations are essential to understand the physical nature of such  structures.}
{ We investigate the kinematics and PM of spectrally separated structures in the FS~Tau~B HH flow. Collating these data makes it possible to understand the origin of these structures and to explain the unusual behavior of the jet. On the other hand, the study of emission profiles in the associated reflection nebulae allows us to consider the source of the outflow both from edge-on and pole-on points of view. }
   {We present the observational results obtained with the 6 m telescope at the Special Astrophysical
Observatory of the Russian Academy of Sciences using the SCORPIO multimode focal reducer with a scanning Fabry-Perot interferometer. Two epochs of the observations of the FS~Tau~B region in H$\alpha$ emission (2001 and 2012) allowed us to measure the PM of the spectrally separated inner structures of the jet.}
   {In addition to already known emission structures in the FS~Tau~B system, we discover new features  in the extended part of the jet and in the counter-jet. Moreover, we reveal a new HH knot in the HH~276 independent outflow system and point out its presumable source. In the terminal working surface of the jet,
structures with different radial velocities  have PMs of  the same value. This result can be interpreted
as the direct observation of bow-shock and Mach disk regions.  

A bar-like structure, located southwest from the source demonstrates zero PM and can be considered as one more example of deflection shock.
An analysis of H$\alpha$  profiles in the reflection nebulae R1\ and R3 indicates the uniqueness of this object, which can be studied in  pole-on and edge-on directions simultaneously.    
 }

   {}


  \authorrunning{Movsessian et al.}

   \maketitle
%
\section{Introduction}
       
Variable star \object{FS~Tau} (Haro~6-5, HBC~383) \citep{haro1953} is located in the Taurus dark cloud on the distance of 140-145 pc \citep{kenyon2008}. It is surrounded by refection nebulae of various sizes and shapes. About 20\arcsec\ to the west from this star, on the axis of bright refection nebula, one of the first Herbig-Haro (HH) jets (HH~157) was discovered \citep{mundt84}. This jet includes various morphological structures and is
narrowing with the distance in an uncommon way \citep{mundt1991}. The star-like knot,
located near the apex of this nebula, was recognized as
the source of the outflow and designated as \object{Haro~6-5~B}
(HBC~381, FS~Tau~B) \citep{mundt84}.

Long-slit spectroscopy revealed the bipolarity of \object{the HH~157} outflow and confirmed the reflection nature of the conical nebula in its apex  \citep{mundt87}. Low absolute values of radial velocities, observed both in the jet and the counter-jet, point to the flow orientation that is close to the plane of the sky  \citep{EM1998}. Further observations with Hubble Space Telescope (HST)  revealed the compact bipolar refection nebula,
similar to \object{HH~30}, in the FS~Tau~B location \citep{ray1996,krist1998}. The source itself is not visible directly because it is located behind the dust lane, which bisects this nearly edge-on oriented nebula \citep{krist1998,padgett,woitas}.
The FS~Tau~B system, especially its circumstellar disk, became the
subject of recent studies in various wavelength ranges \citep{yokogawa2001,kirchschlager2016}.
It is worth noting that in vicinity of FS~Tau~B a bipolar molecular  outflow was also detected \citep{mottram}, which is about 2\arcmin \ in length. Its blueshifted and  redshifted lobes coincide with the jet and
counter-jet parts of FS~Tau~B HH~157 optical outflow, respectively.

The complex inner
morphology of  jets from young stellar objects  was effectively revealed by scanning  Fabry-Perot interferometer (FPI) observations, which can successfully separate structures with high and low radial
velocity \citep{morse1992,morse1993,hartigan2000,mov2007}. This observational technique provides  high spectral resolution data with full two-dimensional coverage in the large field of view. Notwithstanding the small spectral range, which
allows us to
investigate only one emission line, it is a very powerful method to study extended emission objects
such as HH objects and optical jets.

 In fact, only spectro-imagery allowed for the revelation of the morphology of different
kinematical structures not only in the working surfaces but in the inner structures of jets as well
\citep{morse1993, hartigan2000, mov2000, mov2007, mov2009}. 
For a better understanding of the physical nature of such structures it is very important to combine the radial velocity data with proper motions (PM). We developed a new method of PM measurement for jet structures with various radial velocity  using two epochs of scanning
FPI observations. It provides  both the spectral separation of the structures
by their radial velocity and the measurement of their PM. It was already tested on HL~Tau jet, for which we
revealed internal structures with surprisingly different radial
velocities but with the same PM \citep{mov2007,mov2012}.

 The present paper extends this approach to the FS~Tau~B HH outflow system. We focus on the full kinematical picture of various emission structures inside the jet and its working surface. In addition, in the course of our study dramatic changes of H$\alpha$ profiles in the jet source and its reflection nebula have been found; we discuss these changes as well.   

\section{Observations and data reduction}

Observations were carried out in the prime focus of the
6 m telescope of Special Astrophysical Observatory (SAO) of the Russian Academy of Sciences in two epoches: 19 Sept. 2001 and
22 Nov. 2012 in good seeing conditions of about
1\arcsec. We used an FPI placed
in the parallel beam of the SCORPIO focal reducer. This
device is described by \citet{AM2005} and
the SCORPIO capabilities in the scanning FPI observational mode are described by \citet{moisav}. 

The detector used during the first epoch of observations was a TK1024 1024$\times$1024 pixel CCD 
array. Observations were performed with 2$\times$2 pixel binning to reduce the readout time, 
so 512$\times$512 pixel images were obtained in each spectral channel. The field of view 
was 4.80\arcmin\ and the scale 0.56$\arcsec$ per pixel (after binning). 

The scanning interferometer, used in these observations, was Queensgate ET-50. This instrument operates in the 501st order of interference at the H$\alpha$ wavelength and provides a spectral resolution of FWHM$\approx$ 0.8\AA\ (or $\approx$40 km s$^{-1}$) for a range of $\Delta\lambda$=13\AA\ (or $\approx$590 km s$^{-1}$) free from order overlapping. The number of spectral channels was 36 and the size of a single channel was $\Delta\lambda\approx$ 0.36\AA\ ($\approx$16 km s$^{-1}$).
The exposure time was 160 seconds per channel and the total exposure time was 96 min.

During the second epoch observations the detector was an EEV 40-42 2048$\times$2048 pixel CCD array. The observations were performed with 4$\times$4 pixel binning and therefore 512 $\times$ 512 pixel images were obtained in each spectral channel. The field of view was 6.1\arcmin\ for a scale of 0.72\arcsec\ per pixel (after binning). In both epochs an interference filter with FWHM = 15\AA\ centered on the H$\alpha$ line was used for pre-monochromatization \citep{moisav2008}. 

The scanning interferometer was ICOS FPI operating in the 751st order of interference at the H$\alpha$ wavelength, providing spectral resolution of FWHM $\approx$ 0.4\AA\ (or $\approx$20 km s$^{-1}$) for a range of $\Delta\lambda$=8.7\AA\ (or $\approx$390 km s$^{-1}$) free from order overlapping. The number of spectral channels was 40 and the size of a single channel was $\Delta\lambda\approx$ 0.22\AA\ ($\approx$10 km s$^{-1}$).
The exposure time was 200 seconds per channel and the total exposure time was 133 min.

\begin{figure}
\centerline{\includegraphics[width=20pc]{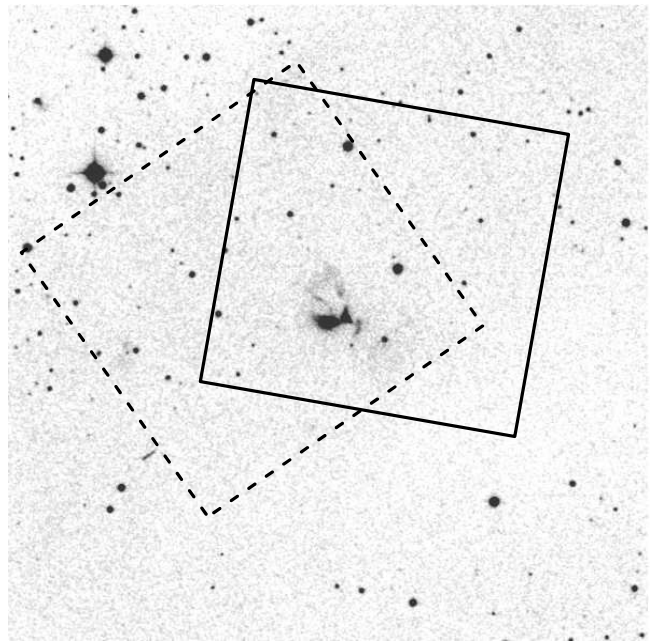}}
\caption{10\arcmin$\times$10\arcmin \ DSS-2 image of the FS Tau B outflow system. Regions, covered by 
FPI, 
are outlined with rectangles, solid for 2001 and dashed for 2012 epochs.}
\label{fig1}
\end{figure}

We reduced our interferometric observations using the software developed at the SAO \citep{moisav,moisav2008} and the ADHOC software package{\footnote{The ADHOC software package was developed by J. Boulesteix (Marseilles Observatory) and is publicly available on the Internet.}}. After primary data reduction, subtraction of night-sky lines, and wavelength calibration, the observational material represents so called data cubes. The corrections of data for telescope guiding errors and for variations of atmospheric transparency and seeing were performed by a method, described by \citet{moisav}, which significantly decreases their influence. Final data cubes were subjected to optimal data filtering, which included Gaussian smoothing over the spectral coordinate with FWHM = 1.5 channels and spatial smoothing by a two-dimensional Gaussian with FWHM = 2--3 pixels.

Using these data cubes, we spectrally separated  the
structures in the outflow system. Then PMs were measured for  the selected structures using observations in both epochs. For the PM estimation a
method of two-image optimal offset computation by
means of cross-correlation was used (realized as  procedure by F. Varosi and included in IDL astronomy library\footnote{https://idlastro.gsfc.nasa.gov/}).

\section{Results and discussion}

\subsection{Overview}

\begin{figure}
\centerline{\includegraphics[width=18pc]{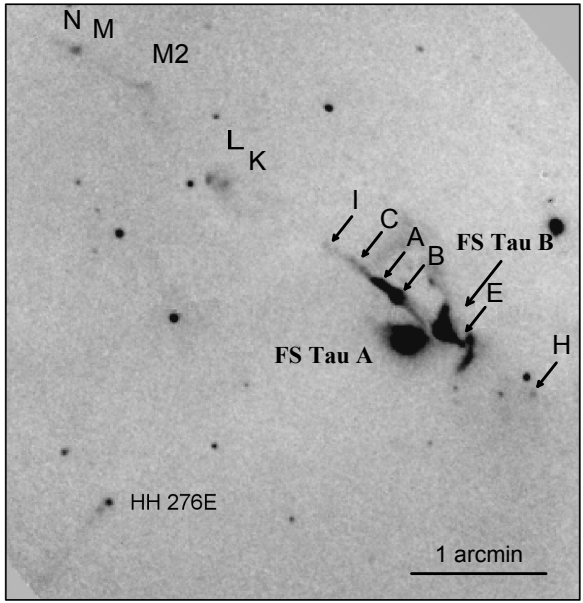}}
\caption{Image of the field, covered by FPI observations in 2012, in the H$\alpha$ line integrated over all channels. Emission structures in the HH~157 flow are denoted using the nomenclature of \citet{EM1998}. Newly discovered knot M2 and Herbig-Haro object HH~278E are shown as well.} 
\label{fig2}
\end{figure}

  In Fig.\,\ref{fig1} the fields covered by SCORPIO\ camera during FPI observations in 2001 and 2012 are outlined on the 10\arcmin$\times$10\arcmin  \ image of DSS-2 red plate. As can be seen, the field, observed in 2012, includes more structures of the outflow system; moreover, it has  a deeper limit, allowing the detection of many faint features. In addition to emission features, the field includes the bright cone-shaped reflection
nebula R1, the star-like reflection nebula R3 near the position of the source ,and the reflection
nebula R2 and cavity walls seen in reflected light \citep[see][Fig.8]{EM1998}. In the outflow
system the central part of the jet as well as the terminal shock region with knots L and K, and the northeastern extension of the flow with knots M and N (nomenclature of \citealt{EM1998}) are covered (Fig.\,\ref{fig2}). Moreover, we discovered an additional linear emission structure connected with a faint bow-shaped wing  designated as M2
between the bow shock and the knot M, which is barely visible on the [\ion{S}{ii}] image of \citet{EM1998}.

\begin{figure}
\centerline{\includegraphics[width=20pc]{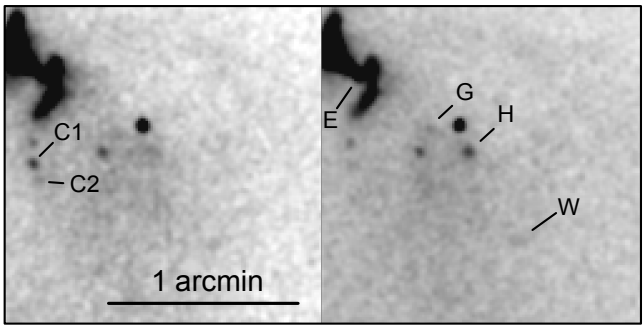}}
\caption{Southwestern side of FS~Tau~B flow. The left panel image corresponds to the radial velocity of +39 km s$^{-1}$ and the right image to + 79 km s$^{-1}$. Emission knots are indicated using the nomenclature of \citet{EM1998}; also the newly discovered knot W is shown.} 
\label{fig3}
\end{figure}

In the counter-jet we detected the emission knots E (near the R3 star-like nebula),  G, and H  as well as the more distant faint knot, which we denote as W. Alongside this linear chain of knots,  C1 and C2 knots (see Fig.\,\ref{fig3}) are also visible, but they probably do not belong to the  FS~Tau~B counter-jet
flow.

The field of view also includes the independent outflow \object{HH~276} \citep{EM1998}. In this system we found
an additional bright knot, designated as HH~276~E, which
traces this flow in the southeastern direction.

We estimate radial velocities and PM for various structures of FS~Tau~B jet, which were visible in both data sets, and compared these with the values presented by \citet{EM1998}. It should be noted that our observations were separated by about 11 years, and this time interval is longer than the cooling time of shocked structures. Indeed we detect some
morphological changes in several knots.
Investigation of the emission lines profiles near the source (R3 reflection nebula) and in the bright reflection nebula R1 is of a special 
interest \citep{movPPVI}.
Below we discuss all these results in detail.  

\subsection{ Morphology of the HH~157 flow inner structures}

\begin{figure*}
\centerline{\includegraphics[width=32pc]{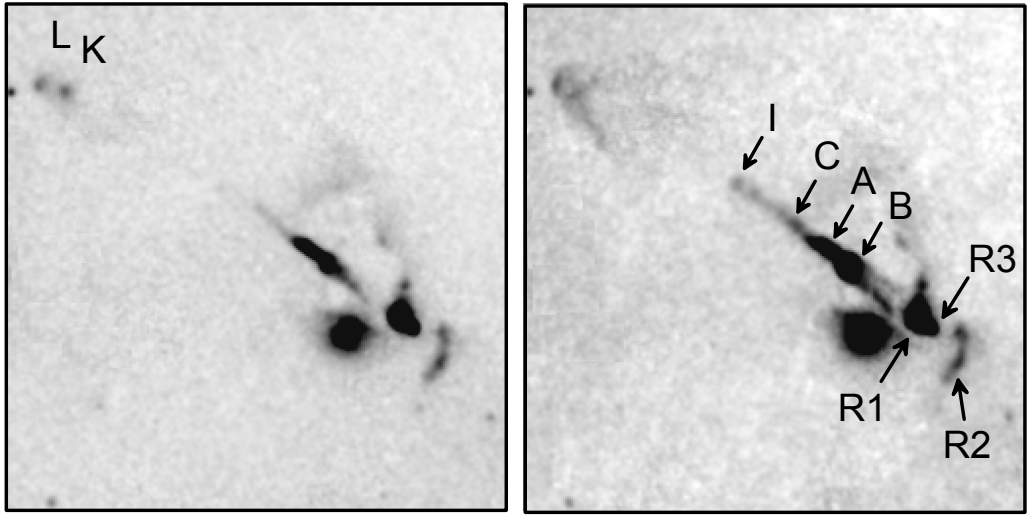}}
\caption{Two channels from the observations with FPI in 2012, which show the variations in the
morphology of FS~Tau~B jet, depending on heliocentric radial velocities; the left panel corresponds to
$-$77 km s$^{-1}$ and the right one to $-$36 km s$^{-1}$.}
\label{fig4}
\end{figure*}

 The appearance of certain parts of the HH~157 flow strongly depends on absolute radial velocity, which can be
seen from the images in various channels obtained during our  2012 observations (Fig.\,\ref{fig4}).
The main jet is wider in low velocity channels compared with the higher velocity channels, similar to the
jet from HL~Tau \citep{mov2007}.
With better clarity these changes can be seen in a channel movie, which is available in the electronic version of the paper.  Considering that the jet consists of a number of small internal working surfaces,  this feature may have two causes.  On the one hand, each internal working surface includes a compact high velocity Mach disk and  low velocity extended bow shock. On the other hand, in each bow-shock radial velocities are maximal near its apex. 

Most prominent feature in the FS~Tau~B jet is the bow-shaped knot B (see Fig.\,\ref{fig2} for details). A linear emission structure extends from this bright
knot toward the source; however, it does not coincide with the axis of the jet. The nature of this structure is
discussed below. The bow-shock region (knots K and L) demonstrates most 
significant changes in
morphology with velocity (Fig.\ref{fig4}).  In high radial velocities both condensations appear as
two compact knots, while in lower velocities they turn into separate bow-shaped structures. More
distant parts of the flow (knots N, M, and newly discovered M2) do not undergo noticeable variations
depending on velocity. This can be explained by the decrease of ambient medium and flow density ratio, which results in a bright
bow shock and a very faint Mach disk \citep{hartigan1989}.

 A narrow counter-jet, ending with relatively bright knot E, is visible near the source in positive
radial velocities while jet as whole has negative velocity\textbf{}. The counter-jet points in
the direction of knot H. It should be noted that the counter-jet, knots G, H, and the newly discovered
knot W, are better seen in the channels near +80 km s$^{-1}$ velocity, but knots C1 and C2 differ by much
lower velocity (+40 km s$^{-1}$), and their  relation to counterflow is doubtful (see Fig.\,\ref{fig3}).

Finally, the unusual appearance of the FS~Tau~B jet, which narrows along with the distance from the source,
can be a result of exceptionally bright and extended bow-shaped knot B. Thus, it could be not associated with the variations in the flow
collimation.

\subsection{Radial velocities }

\begin{figure}
\centerline{\includegraphics[width=16pc]{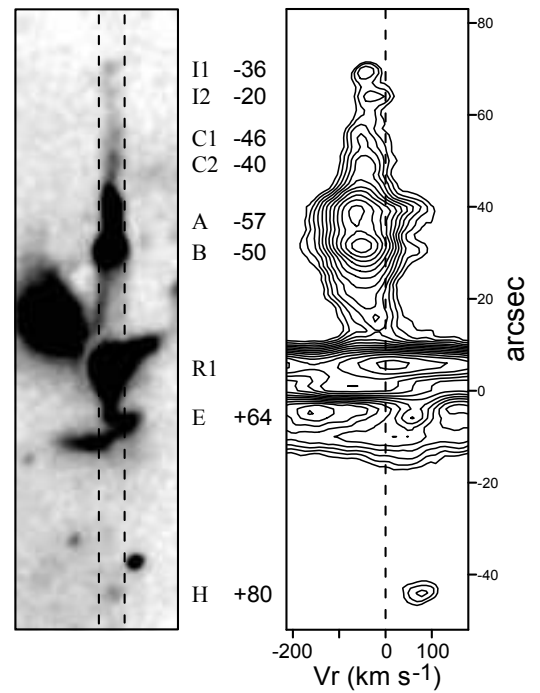}}
\caption{Monochromatic H$\alpha$ image of the FS~Tau~B jet and the orientation of pseudo-slit (left) and corresponding PV diagram. Zero velocity is shown by a dashed line (right). Knots and their radial velocities are indicated between the panels. }
\label{fig5}
\end{figure}

Because of higher spectral resolution, we concentrate on the radial velocities, obtained in the
second epoch; in any case, the differences of radial velocities between two epoch observations do
not exceed 5 km s$^{-1}$ for knot B and are lower than 2 km s$^{-1}$ for other knots. To compare our results with the previous measurements, obtained by means of long-slit spectroscopy
\citep{EM1998}, we  generated a pseudo-slit, which is  oriented along the axis of the jet on the base of the data cube obtained in
2012. Fig.\,\ref{fig5} shows a
jet image in H$\alpha$ emission and position-velocity (PV) diagram. This
diagram only covers the central part of the jet near the source and does not include bow shock because
the jet axis does not point toward the bow apex. Intensities in the PV diagram are presented in the
logarithmic scale because of their strong variations along the jet.

From the PV diagram it is obvious that the absolute radial velocity increases from the source up
to the knots B and A. Beyond this bright central part of the jet and up to the knot I radial velocity
exhibits  low changes. There are evident signs of the existence of low velocity wings in H$\alpha$
profiles in the positions of knot B and especially knot A.

Our data are in good accordance  with measurements of \citet{EM1998},   especially for the brightest knots, confirming the low absolute values of radial velocities in the jet
and counter-jet. The disagreement between old and new data,\ however, increases toward the end of jet. The probable reason is that the jet is curved and during the observations of \citet{EM1998} the long slit of spectrograph missed its end.   

\begin{figure}
\centerline{\includegraphics[width=20pc]{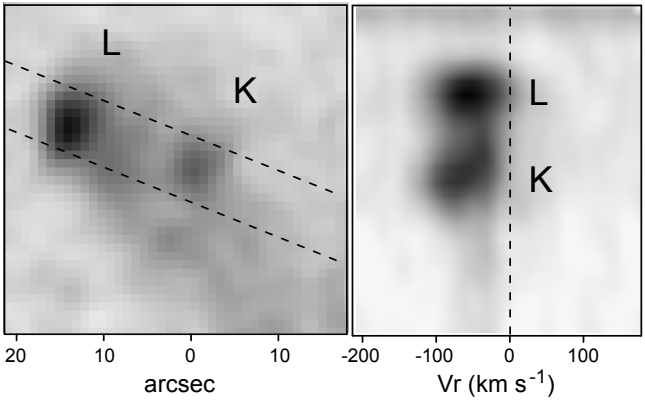}}
\caption{Monochromatic image of the FS~Tau~B jet bow-shock part and pseudo-slit orientation (left) and corresponding PV diagram (right). North is up.}
\label{fig6}
\end{figure}

Therefore, we slightly changed the orientation of pseudo-slit to  extend the intensity map and PV diagram of HH~157 flow to its bow-shock region, defined by knots K and L. Results are shown in Fig.\,\ref{fig6}, which is similar to  Fig.\,\ref{fig5}. The existence of two velocity
components in the knot K is obvious. As we already discussed in the previous section, the low velocity
component represents bow-shaped structure, while high velocity component is produced in the compact knot.
In knot L the split in velocity is not so evident but the low velocity wing is clearly visible.
Full data about heliocentric radial velocities of the knots, indicated on the PV diagram and in the extended parts of the
flow, are gathered in Table\ \ref{table}. 

\begin{table*}
\caption{Proper motions and radial velocities of knots in the FS~Tau~B outflow}
\label{table}
\centering
\begin{tabular}{l c c c c c}
\hline\hline
Knot & Distance\tablefootmark{a}  & V$_{tan}$\tablefootmark{b} & V$_{tan}$ & PA & V$_{r}$ \\  & (arcsec) & (km s$^{-1}$)
&   (arcsec yr$^{-1}$) & (deg) & (km s $^{-1}$) \ \ \ \ \\
\hline
B               &  31.5 &   395 $\pm$ 30   &  0.62 & 46   &   $-$50 \\
A               &  38.7 &   231 $\pm$ 35 & 0.37  &   41   &   $-$57 \\
C2              &  48.5 &    -   &   -    &  - & $-$40 \\
C1              &  53.9 &   248 $\pm$ 25  & 0.39  &   55   &   $-$46    \\
I2              &  63.5 &    -   &   -    & - & $-$20 \\
I1              &  68.5  &   279 $\pm$ 22 &  0.44  &   56   &   $-$36 \\
K$_{high~vel}$  &  120.2   &   360 $\pm$ 23 & 0.55 &   55 & $-$77 \\
K$_{low~vel}$   &  120.8     &   -    &   -    & - &  $-$30 \\
L$_{high~vel}$  &  127.7     &   285 $\pm$ 25 & 0.42 &   58   &   $-$60 \\
L$_{low~vel}$       &  128.8    &   290 $\pm$ 36 & 0.46  &   56   &   $-$30 \\
M2              &  176     & -  & -  &   - &    $-$30            \\
M               &  209  &    -  & -  &   - &    $-$45           \\
N               &  220  &    -  & -  &   - &    $-$45    \\
E               &  5.9  &   180  $\pm$ 26 & 0.29  &   220  &   +65  \\
G               &  31.7 &    -   &    -  & - &  +78 \\
H               &  44.7 &   160 $\pm$ 20 &  0.25  &   230  &   +80  \\
\hline
\newline
\end{tabular}
\tablefoot{
\tablefoottext{a}{Distances of knots are measured on 2012 image, taking the R3 bright knot as a zero point.}
\tablefoottext{b}{Tangential velocities correspond to 150 pc as a distance of the flow.}
}
\end{table*}

\subsection{Proper motions}

We measured PM for all the jet structures that were detectable in the images of both epochs. The results are presented in Table\ \ref{table}: i.e., distances for each knot, measured from the center of R3 reflection nebula; values of tangential velocities, computed for the distance of 150 pc for easy comparison with the data of \citet{EM1998}; PAs; and radial velocities.

The high values of PM of knots in the FS~Tau~B jet were already known from previous studies; nevertheless, the
data presented in this work   are of additional  interest because they refer to the knots selected in the
same velocity channels.
Moreover, in the cases in which inner structures undergo strong morphological changes depending on the radial velocity,
we present PM values for these features in both velocity channels.
To achieve  a better comparison of morphological details  in these cases, we performed the rebinning of the second epoch data cube to bring both of them to the same velocity steps. In general our PM measurements are in good accordance with the values presented by
\citet{EM1998}.

\begin{figure*}
\centerline{\includegraphics[width=36pc]{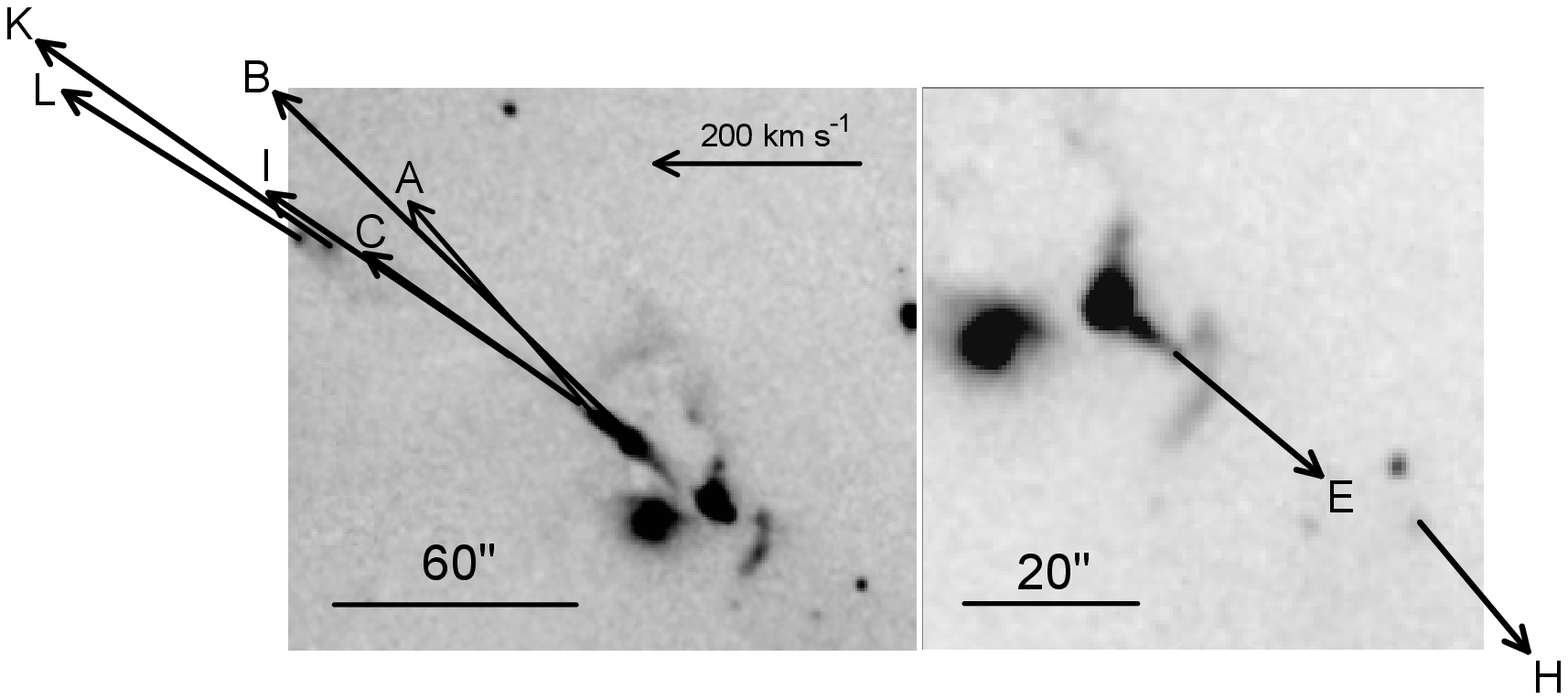}}
\caption{Monochromatic images of the FS Tau B jet and the counter-jet with superposed PM vectors of individual knots.
Images are obtained by integrating over the blueshifted (left) and redshifted (right) components of
H$\alpha$ emission.}
\label{fig7}
\end{figure*}

\begin{figure}
\centerline{\includegraphics[width=10pc]{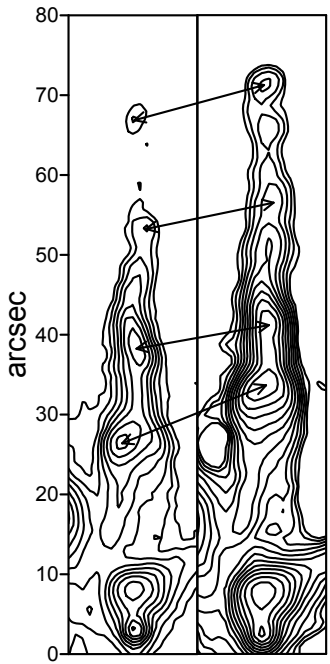}}
\caption{Monochromatic images of FS Tau B jet in 2001 (left) and 2012 (right). Arrows connect the same knots in both epochs.}
\label{fig8}
\end{figure}

\begin{figure}
\centerline{\includegraphics[width=16pc]{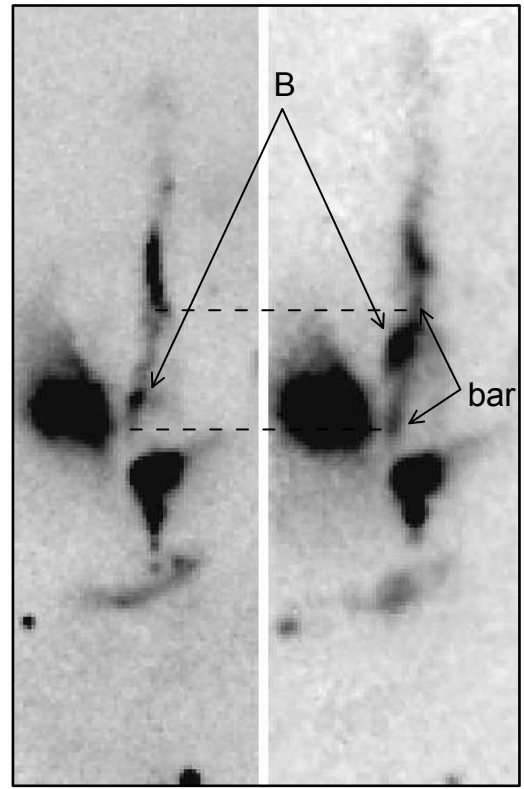}}
\caption{H$\alpha$ image of the FS Tau  B jet obtained by \citet{EM1998} in 01.01.1987 (left) and monochromatic image of the FS Tau B jet in low radial velocity channel ($-$15 km s$^{-1}$) obtained in
19.09.2001 (right). Fast-moving knot B and the stationary bar are denoted.}
\label{fig9}
\end{figure}

The PM vectors  for the various structures in the HH~157 jet and the counter-jet are shown separately in Fig.\,\ref{fig7}.
To achieve better representability the images in both frames of Fig.\,\ref{fig7} were  obtained by
integration of channels, corresponding to  blueshifted and redshifted ranges of H$\alpha$
emission.

Fig.\,\ref{fig8} shows isolines of integrated  H$\alpha$ images of FS~Tau~B outflow for both
epochs; the shifts of    selected knots are indicated by arrows. The high PM of knot B is obvious. It
results in the striking ``sliding'' of knot B along the jet. Knot I2,  which is barely visible on
the 2001 image and was not detected at all in the images of \citet{EM1998}, is well visible in
2012.

 As was described previously,    the working surface of the jet contains  two separate structures
designated as knots K and L. Their morphology strongly changes with radial velocity. We tried to
measure PM of high and low velocity structures in these knots separately. But the detection limit in 2001 observations was lower, which makes the low velocity structure in K knot blurred and poorly pronounced.
Therefore its PM was not possible to measure with sufficient precision. In comparison, in knot L both high and low velocity
structures are clearly outlined and measurable. PM of these velocity structures are the same, although the
radial velocities are very different. This result is consistent with  analogous   measurements of knots in the \object{HL~Tau} jet \citep{mov2012}.

Different radial velocity structures in the knots of the jets and terminal working surfaces
are usually interpreted as two principal shocks regions: a
``reverse shock (Mach disk)'', which decelerates the supersonic flow, and a ``forward shock'',
which accelerates the material with which it collides \citep{hartigan1989}.
The morphologies of these structures are very different:  reverse shock appears as a compact knot
while forward shock is represented by a bow-shaped extended structure.

Taking this approach into account, it is possible to obtain
more precise estimations for the orientation angle of flows. Since in the reverse shock
region the ratio $\zeta$ between the pattern speed of the knots and the flow speed of the jet
particles is equal to one (i.e., $v_{pattern} = v_{flow}$), the orientation angle can be calculated by
the ratio of $v_{rad}$ to $v_{tan}$ \citep{EM1992}.
To apply this method for the FS\ Tau B flow,  we chose knot L, for which both shock regions are clearly divided. Using tangential and radial velocities of the
high velocity structure of the knot
L, we found that its inclination to the plane of the sky and its  spatial velocity are equal to 12$^\circ$ and 300 km s$^{-1}$, respectively.
On the other hand, these values should vary with the distance from the source due to precession. 

Another interesting feature in the FS~Tau~B flow is the bar-shaped structure  with length about 20$\arcsec$ mentioned above; this structure starts at a point located to the northeast from the
source. In 2012 this structure is a little disturbed by the fast-moving B knot, but
the structure is clearly visible in 2001. Therefore, we demonstrate its appearance in 1987 on the H$\alpha$ image of FS~Tau~B, provided to us by J. Eisl{\"o}ffel,  and  in the monochromatic image obtained with FP etalon in
2001 (Fig.\ref{fig9}). It can be easy concluded that the bar-shaped structure  appears
stationary and, as in the case of \object{HH~47}, can by  interpreted  as a deflection shock region
\citep{hartigan2005,hartigan2011}.

Finally, the unusual appearance of the FS\ Tau B jet, which narrows with the distance from the source,
could be a result of the exceptionally bright and extended bow-shaped knot B. Thus, the narrowing of the jet is probably not related to the variations in the flow
collimation \citep{mundt1991}.

\subsection{Source and reflection nebula R1}

Our data cubes allow us to study the H$\alpha$ emission profiles not only in the spectra of the star
and of its directed outflow, but also in the various reflection nebulae associated with this system.
To be more precise, we must note that the star FS~Tau~B itself is totally obscured, as was shown by
\citet{Stapelfeldt97}, and its optical radiation cannot be observed directly. The bright 
stellar-like knot (R3), located to the northeast from the source, probably represents the upper surface of the circumstellar disk seen edge-on.
Further observations with Wide Field Planetary Camera 2 on the HST revealed the structure of the
source in more detail. Images show that the source of a bipolar jet appears to illuminate a
compact, bipolar nebula which is assumed to be an edge-on protostellar disk similar to HH~30
\citep{ray1996, krist1998}. 
\begin{figure}
\centerline{\includegraphics[width=16pc]{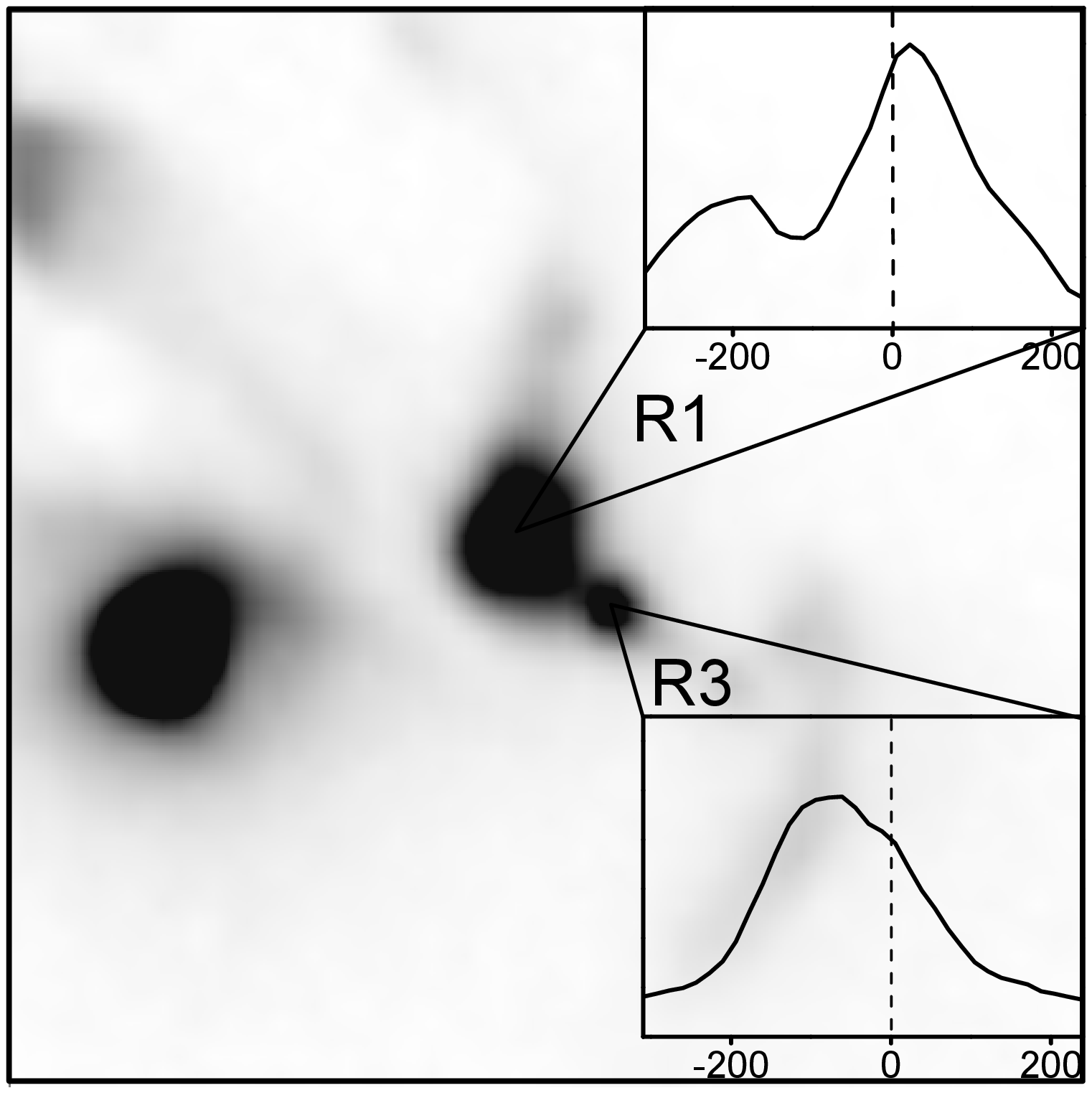}}
\caption{Integrated H$\alpha$ profiles observed in R1 and R3 reflection nebulae. Radial velocities are indicated in the boxes with profiles.}
\label{fig10}
\end{figure}

To study wide H$\alpha$ line
profiles in the reflection nebulae we used first epoch observations, which covered wider
spectral range. As can be seen from Fig.\,\ref{fig10} and the PV diagram (Fig.\,\ref{fig5}), the H$\alpha$ line
profile, which corresponds to R3 reflection nebula, shows broad emission.
In contrast, the averaged profile in the R1 nebula is split into two components.
It should be noted, that the emission profile in the position of R3 nebula is typical for the edge-on T~Tau stars \citep{Appenzeller} and the profile in the R1 nebula is similar to the majority of classical T~Tau stars. In our opinion, reflection nebula R1 works as a mirror via which we observe the light from the source from pole-on point of view; this contrasts with R3 nebula through which we observe an   edge-on system. The situation is somewhat similar to the geometry in the R Mon+NGC 2261
system \citep{JonesHerbig}.

The profile of H$\alpha$ emission, observed in R1 reflection nebula, consists of two peaks: first peak is a wide triangular-shape line, little shifted in the red range and the  second peak is a strongly blueshifted emission with radial velocity of about $-$200 km s$^{-1}$. The first emission peak can be explained by assuming magnetospheric accretion with high accretion rate of about 10$^{-7}$ M$_{\sun}$ yr$^{-1}$ \citep{kurosawa}. The division of emission profile could be produced by absorption in disk wind \citep{kurosawa}, but in this specific case the stellar emission (i.e., first peak) is symmetric and the velocity of blueshifted second peak is too high in comparison with the model. In our opinion, the blueshifted peak  is created by high velocity outflow that is observed close to its axis (via the point of view in R1 nebula). 

The situation completely  differs in R3 nebula. Considering the full H$\alpha$  profile, we can see that it consists of two  blended emission components, one of which       has nearly zero velocity and the other is blueshifted to about $-$70 km s$^{-1}$; it should be noted that this shift is very close to the velocity of the jet in FS\ Tau B flow). We suggest that zero velocity component has the same origin as in R1 nebula (but is more faint) and probably corresponds to stellar emission, obscured by circumstellar disk or torus in contrast with the more extended emission of the outflow. On the other hand, the width of blueshifted emission line exceeds the width of the jet emission much more. We suggest that this component corresponds to the base of the jet where the opening angle of flow is wider in comparison with the jet itself. Such large values for the initial opening angle of several HH jets were estimated from high spatial resolution images obtained with HST \citep{ray1996}. 

It is necessary to note that these assumptions consider a simple single light scattering on the dust in the reflection nebulae.  More complicated effects, such as variable extinction or circumstellar shadowing, are not taken into account. They probably can change the relative intensities of the emission components.

\subsection{HH 276 outflow system\\ }

This independent outflow was discovered by \citet{EM1998} as a diffuse emission tail, almost
perpendicular to FS~Tau jet. The length of this flow is about 150$\arcsec$ and includes several
faint knots. The source of this flow was not known, but an assumption was made that it should be located in northwestern
direction. It was preliminary designated as FS~Tau~C \citep{EM1998}.

As mentioned above, we found a bright emission object in southeastern part of the
studied field. This object consists of two knots and has a faint tail that is well aligned with the axis of HH~276
flow. This object, with coordinates RA = $04^{h}\ 22^{m}\ 12^{s}$, Dec$ = +26\degr\ 56\arcmin\ 19\arcsec$(2000), definitely comprises a part of HH~276 flow, and we designated the object as HH~276E.
Thus, the whole length of HH~276 outflow system extends to about 200$\arcsec$. The radial velocity of this
outflow is about +30 km s$^{-1}$ and undergoes only small changes along the flow. The low values
of radial velocities point to the flow propagation near the plane of sky, just as in
the case of FS~Tau~B jet.

\begin{figure}
\centerline{\includegraphics[width=20pc]{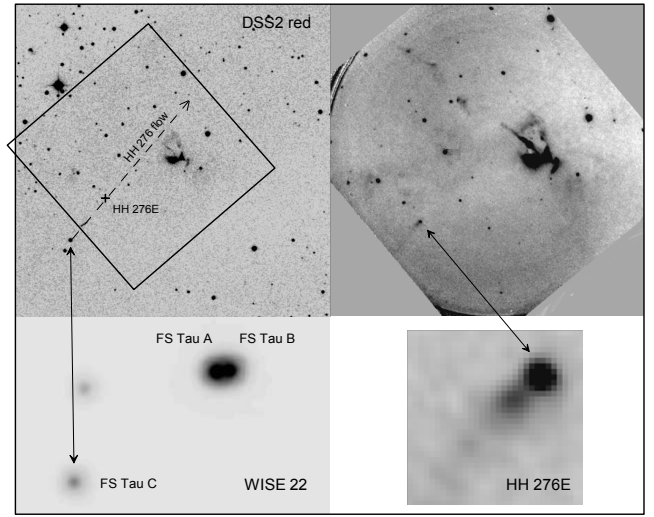}}
\caption{ Positions of  the newly discovered HH 276E object and the proposed source of HH 276 flow (FS\ Tau C). The top left panel shows their location on the DSS-2 red image and bottom left shows the  WISE survey 22 mkm image. The top right panel repeats the image, presented in Fig.\,\ref{fig2}, and the bottom right   shows the enlarged image of HH 276E.     }
\label{fig11}
\end{figure}
        
 Our attention was drawn by the star that lies just near the axis of the flow, namely \object{IRAS~F04192+2647} (CFHT-Tau21). This source was classified as a very low-mass star with undoubted outflow
signatures, such as  near-infrared excess radiation arising from the  accretion
disk and forbidden lines in the spectrum \citep{guieu}. On the other hand,
this source is one of brightest objects in the WISE survey 22 $\mu$m map  of this region.
   Our
long-slit spectroscopy of this source, carried out at Byurakan Observatory, allowed us to
detect \ion{an Li}{i} absorbtion line, which is a further indication of  youth of this star.

We propose this star as a probable source of the HH~276 outflow. This study is  described in more detail in a separate paper \citep{movsessian2019}. 

\section{ Conclusions    }

We have performed two epoch FPI observations of FS\ Tau A/B region in H$\alpha$ emission,
which allowed us to measure the PM of the spectrally separated emission structures in the  FS Tau B HH
outflow system. 

This HH outflow not only narrows with the distance  \citep{mundt1991}, but also
demonstrates the strong variations of the width depending on radial velocity. Namely, the optical flow is more
narrow in the high radial velocity  than in the low velocity channels. In the more distant part of flow, strong variations of morphology are
not observed. Such behaviour of the optical flow can be explained by a stronger contribution of oblique shock
zones in the low velocity channels.

With the aid of the more deep second epoch data we discovered the new emission structures in the extended part of the
jet (knot M2 with the trailing faint
bow-shaped structure) and in the counter-jet (knot W).
On the other hand, a new relatively bright HH object was discovered in the southeastern direction
from FS Tau A/B system. This object
lies on the axis of HH 276 flow and is denoted as HH 276E.
As a presumable source of this independent flow, the star CFHT-Tau21 with undoubted outflow
signatures, described by \citet{guieu}, can be considered.

The PMs of the spectrally separated structures were estimated. An especially high PM
was detected in the bright bow-shaped knot B. We can conclude  that the velocity of the outflow undergoes strong variations
in time.

The terminal working surface of the jet
undergoes apparent morphological variations depending on radial velocity. Two compact knots (K and
L) were detected in high radial velocities, accompanied by two leading bow-shaped low velocity structures. In the case
of knot L the structures with different radial velocities  have PMs of  the same value. This result can
be interpreted
as the direct observation of bow-shock and Mach disk regions in the terminal working surface. These data lead to the  more precise estimations of the inclination to the plane of the sky and the  spatial velocity of knot L, which are equal to 12$^\circ$ and 300 km s$^{-1}$, respectively. On the other hand,
significantly dissimilar PMs of knots K and L suggest their interpretation as  two separate working surfaces,
perhaps created by two independent outbursts.

Aside from moving
knots and bright bow-shaped structures, an unusual bar-like structure, which demonstrates zero PM,
exists in the jet and is located to the southwest from the source. It can be yet another example of the
deflection shock \citep{hartigan2011}, which occurs when high velocity
wind impacts the edge of the cavity
formed by outflow.

Analyses of H$\alpha$  profiles in the R1 reflection nebula and in the point near the invisible
source position (R3) reveals
their strong difference.  Near the source,
where we observe the scattered light in the circumstellar disk, the broad emission was detected,
while the averaged profile in the R1 nebula splits into two components. A reflection nebula probably works as
a mirror
and we observe the same object from the edge-on and the pole-on points of view simultaneously, which
makes FS~Tau~B a unique object.
The large
width of H$\alpha$ line in the position of R3 reflection nebula, in comparison with the emission in
FS~Tau~B jet, can be the result of a wide opening angle of the jet base.

All these results once again demonstrate that the observations with FPI  represent a very powerful method
for the investigation of extended emission
objects, especially if combined with the measurement of the PMs of spectrally separated structures.

\begin{acknowledgements}
 The authors thank the referee for helpful suggestions. We are grateful to Dr. V.A. Moiseev for very important contributions in observations and data reduction. The authors also thank Dr. J. Eisl{\"o}ffel for providing his observational data and helpful discussions. This work was supported by the RA MES
State Committee of Science, in the frames of the research projects number 15T-1C176 and 18T-1C-329.
\end{acknowledgements}
   

\end{document}